# Flux rope merging and the structure of switchbacks in the solar wind

O. Agapitov[1,2], J. F. Drake[3,4], , M. Swisdak[4], S. D. Bale[2], T. S. Horbury[5], J. C. Kasper[6,7], R. J. MacDowall[8], F. S. Mozer[2], T. D. Phan[2], M. Pulupa[2], N.E.Raouafi[9], and M. Velli[10]

Abstract

A major discovery of Parker Solar Probe (PSP) was the presence of large numbers of localized increases in the radial solar wind speed and associated sharp deflections of the magnetic field - switchbacks (SB). A possible generation mechanism of SBs is through magnetic reconnection between open and closed magnetic flux near the solar surface, termed interchange reconnection, that leads to the ejection of flux ropes (FR) into the solar wind. Observations also suggest that SBs undergo merging, consistent with a FR picture of these structures. The role of FR merging in controlling the structure of SBs in the solar wind is explored through direct observations, analytic analysis, and numerical simulations. Analytic analysis reveals key features of the structure of FRs and their scaling with heliocentric distance $R$ that are consistent with observations and demonstrate the critical role of merging in controlling the structure of SBs. FR merging is shown to energetically favor reductions in the strength of the wrapping magnetic field and the elongation of SBs. A further consequence is the resulting dominance of the axial magnetic field within SBs that leads to the observed characteristic sharp rotation of the magnetic field into the axial direction at the SB boundary. Finally, the radial scaling of the SB area in the FR model suggests that the observational probability of SB identification should be insensitive to $R$, which is consistent with the most recent statistical analysis of SB observations from PSP.

[1] Corresponding author agapitov@ssl.berkeley.edu
[2] Space Sciences Laboratory, University of California Berkeley, Berkeley, CA, USA
[3] Department of Physics, the Institute for Physical Science and Technology and the Joint Space Institute, University of Maryland
[4] Institute for Research in Electronics and Applied Physics, University of Maryland, College Park, MD
[5] The Blackett Laboratory, Imperial College London, London, UK
[6] BWX Technologies, Inc., Washington DC
[7] Climate and Space Sciences and Engineering, University of Michigan, Ann Arbor, MI
[8] Code 695 NASA Goddard Space Flight Center, Greenbelt, MD
[9] Johns Hopkins Applied Physics Laboratory
[10] Department of Earth, Planetary and Space Sciences, University of California, Los Angeles, CA

## 1. Introduction

A recent major discovery of Parker Solar Probe (PSP, Fox et al., 2016) was the presence of large numbers of localized velocity spikes associated with magnetic structures containing sudden deflections in the local radial magnetic field at 35.7-50 solar radii (RS) near the first PSP perihelion (Bale et al. 2019; Kasper et al. 2019, and others). The observed rotation angle inside these structures varies up to full reversal of the radial magnetic field component (Dudok de Wit et al. 2020), hence inspiring their designation as "switchbacks" (SB). The time duration of a SB from the PSP data varies over a wide range from tens of seconds to tens of minutes (Dudok de Wit et al. 2020). Proton temperature enhancements are often associated with SBs (Agapitov et al. 2020; Krasnoselskikh et al. 2020; Larosa et al. 2021; Woodham et al. 2020). The plasma temperature increase inside SBs suggests that they may be magnetically isolated from the ambient solar wind. The constancy of the electron strahl pitch angle across the switchback (Kasper et al 2019) is an important constraint on SB generation mechanisms. Localized reversals of the magnetic field were observed in the coronal hole plasma of the Ulysses polar•crossing data set (e.g.,(Balogh et al. 1999; Borovsky 2016; Neugebauer & Goldstein 2013; Yamauchi et al. 2004) , and were also seen in the coronal hole plasma at 1•AU (e.g., Kahler et al. 1996) and at 0.3•AU (Horbury et al. 2018). However, PSP measurements from closer to the Sun (PSP's first perihelion was at 35.7 RS or 0.174 AU whereas Helios A and Helios B had perihelia at 0.31 and 0.29 AU) have revealed that SBs undergo significant evolution as they propagate outward from the sun from 30 to 50 RS – the distance covered by measurements in the first six encounters (the first three encounters had perihelia at 35.7 RS). Compared to the SBs at perihelion, the SBs at 50 RS are more relaxed structures (Mozer et al., 2020): (1) the temperature of plasma inside the SB is reduced so that the difference with the ambient solar wind plasma temperature becomes insignificant; (2) the wave activity inside SBs and on their boundaries decreases by a factor of 5-10; (3) the rotation angle of the magnetic field direction inside SBs increases by a factor of two; and (4) SB boundaries more closely resemble properties of more stable tangential discontinuities (Akhavan-Tafti et al. 2021). Thus, compared to the study by (Horbury et al. 2018) in which there was little variation between SBs from Helios at 0.3AU, Wind at 1AU, and Ulysses at 2.4AU, it appears that SBs are evolving between 30 and 50RS. The radial evolution is towards a reduction in the density and temperature jumps across switchback boundaries.

Generation theories for SBs focus either on processes occurring deep in the solar atmosphere or in the expanding solar wind. One possibility is that SBs are produced locally through the amplification of turbulence in the expanding solar wind (Martinović et al. 2020; Shoda et al. 2021; Tenerani et al. 2020) or shear-driven turbulence (Landi et al. 2006; Ruffolo et al. 2020; Schwadron & McComas 2021). However, a key observation -- the sharp rise in the ion temperature at the boundaries of the switchback (Farrell et al. 2020; Larosa et al. 2021; Mozer et al. 2020) may be inconsistent with a model based on the amplification of Alfvénic turbulence.

Coronal sources (Dudok de Wit et al. 2020; Krasnoselskikh et al. 2020; Macneil et al. 2020; Woodham et al. 2020) that include reconnection between open and closed magnetic flux (interchange reconnection) (Drake et al. 2021; Fisk & Kasper 2020; Zank et al. 2020) or reconnection associated with jets (Dudok de Wit et al. 2020; He et al. 2020; Sterling & Moore 2020) can produce magnetic flux ropes (FRs) and inject them into the solar wind. Small-scale magnetic FRs in the solar wind at 1 AU were reported by Moldwin et al., (1999) from IMP 8 and WIND spacecraft observations. Magnetic reconnection as the source for flux ropes in the Earth magnetosphere (magnetic flux transfer events at the Earth's magnetopause) has been previously discussed (Russell and Elphic, 1978; Lee and Fu, 1985; Slavin et al., 2003). It was suggested that FRs in the Earth magnetosheath (flux transfer events) resulted from the ripping off of flux tubes from the near-tail dayside magnetopause through magnetic reconnection based on ISEE 1 and ISEE 2 (Russell and Elphic, 1978; Lee and Fu, 1985) and GEOTAIL (Slavin et al., 2003) observations. The structures were observed to be force-free FRs without significant velocity enhancement and with comparable perturbation of all magnetic field components and similar radial and transverse spatial scales. The statistical properties of FR structures in the solar wind were reported by Chen et al.( 2020, 2021) based on events recorded during PSP's first approach to the sun. Drake et al. (2021) used two-dimensional particle-in-cell simulations to study the hypothesis that SBs are flux-rope structures that are ejected into the solar wind by bursty interchange reconnection. It was found that FRs with radial-field deflection (up to full reversal), nearly constant B magnitude, and temperature enhancements are naturally generated by interchange reconnection; and, FR initial conditions relax into structures that match PSP observations reasonably well (Drake et al. 2021). The possible connection between SB's and FR's was discussed by Chen and Hu (2021). Chen et al. (2021) showed that flux ropes can be embedded within switchbacks.

The open question is the physical processes transforming FRs produced during interchange reconnection deep in the corona into the flux ropes that characterize SBs. Compared with FR's expected from interchange reconnection, SB's: have axial magnetic fields that are strong compared with the magnetic field that wraps the axial field; are highly elongated along the direction of the ambient solar wind magnetic field; and are characterized by flows with high Alfvenicity. A process that can play a key role in the evolution of SBs is magnetic reconnection on the boundaries (Phan et al. 2020). Features found at the boundaries of several SBs (Froment et al. 2021) indicate that reconnection with the ambient solar wind field can play a role in the erosive decay of SBs. However, observations of switchbacks in the entire range of heliocentric distances from 20 RS to 2.4AU suggests that reconnection with the solar wind magnetic field is suppressed, presumably due either to the velocity shear (Chen et al. 1997; Dahlburg et al. 1997) or diamagnetic stabilization (Phan et al. 2010, 2013; Swisdak et al. 2010). In addition to reconnection with the ambient solar wind magnetic field, FRs injected into the solar wind via interchange reconnection can also undergo merger (Drake et al., 2021). The FR (magnetic island) coalescence process has been studied by numerical simulations (Pritchett, 2008; Oka et al., 2010; Odstrcil et al., 2003; Zhou et al., 2014), by remote spacecraft measurements of electrons accelerated during merging process (Song et al., 2012), direct measurements during crossing of a series of merging flux ropes of CME in the solar wind, and by *in situ* measurements by the four MMS spacecraft at the terrestrial magnetopause (Zhou et al., 2017). (Drake et al., 2006; Pritchett, 2008; Oka et al., 2010; Song et al., 2012; Zhou et al., 2014) find that merging is very dynamic and releases large amounts of energy. The comprehensive numerical study of FR coalescence in guide field reconnection by Zhou et al. (2014) showed that the coalescence of macroscopic FRs can provide significant energy dissipation and can be an efficient mechanism for particle energization. Flux rope merging was active in the numerical model of switchback formation presented by Drake at al., (2021) where a train of FRs merged through reconnection.

In this paper we explore the structure of FRs sourced from interchange reconnection in the solar corona as they propagate outward in the solar wind, including the scaling of their cross-sectional area (in the plane transverse to the SB axis - often this plane is close to the R-N plane in heliospheric coordinates, which is used to present the cross-sectional plane in the following for simplicity of notation), their aspect ratio (R versus N direction), their interaction during propagation, the energetics of merging, and its consequences for evolution of the FR structure. We demonstrate that the outward expansion of FR's in the solar wind combined with flux rope merging causes FR's generated during interchange reconnection to transition to FR's that match the character of SB's. To perform this study we present theoretical arguments, the results of particle-in-cell (PIC) simulations based on the further development of the model presented by Drake et al. (2021) and PSP observations.

## 2. Switchback characteristics from PSP measurements

We use measurements from PSP of electric and magnetic fields made by the PSP FIELDS suite of instruments (Bale et al. 2016). The vector magnetic field is measured from DC to several tens of Hz by the fluxgate magnetometer (MAG) while magnetic fluctuations above 10 Hz are measured by the Search-Coil Magnetometer (SCM, Jannet et al. 2021). The DC electric measurements are made by the EFI electric antennas. All these data products are provided by the Digital Fields Board (DFB, Malaspina et al. 2016). The sampling rate of the waveforms corresponds to the survey cadence during the early part of the solar encounter phase. During the close encounter phase this cadence increases fourfold. The proton velocity, density and temperature are provided by the SWEAP suite (Kasper et al. 2016). SPC Faraday cups (Case et al. 2020) provide moments of the reduced distribution function of ions: density, velocity and radial component of the thermal velocity. Their cadence is 0.22 seconds. Finally, we consider the electron pitch angle distribution from the Solar Probe ANalyzer-Electron (SPAN-E, Whittlesey et al. 2020), whose cadence is 28 seconds.

A typical switchback is a perturbation of the solar wind structure containing a proton bulk velocity spike and an associated localized deflection of the magnetic field direction. The magnetic field structure of a SB recorded at about 36 RS from the Sun (November 5, 2018 – the first PSP perihelion) is shown in Figure 1a (the components are shown in the RTN coordinate system with R the radial direction directed from the Sun center, N the normal to the ecliptic plane component, and T the azimuthal component). The sharp rotation of the direction of the magnetic field at the boundary while remaining nearly constant in magnitude, and the radial magnetic field changing sign, are typical characteristics of these events. The boundaries range in widths from tens of km (several proton inertial lengths) to tens of thousands of km (Krasnoselskikh et al. 2020; Larosa et al. 2021; Mozer et al. 2020). The perturbation of the proton bulk velocity (Figure 1b) follows the magnetic field perturbation illustrating the Alfvénicity of SBs, i.e. $\Delta \vec{B}_{SB} \sim \Delta \vec{V}_{SB}$. The plasma density enhancements

(highlighted by light blue in Figure 1c) are typical for SB boundaries (~30% on average (Farrell et al. 2020)). The enhancement of the parallel proton thermal velocity inside the SB to 63±3 km/s (with the ambient value of 55±3 km/s) is shown in Figure 1d. Switchbacks often have a complex internal structure highlighted in Figure 1 with dark red – the structure of this particular switchback has been resolved making use of the Grad-Shafranov reconstruction by Chen and Hu (2021) and showed that this switchback consists of three flux ropes confirming the presumption for this event by Drake et al. (2021).

A characteristic parameter for SBs is the angle that the axial magnetic field makes with respect to the direction of solar wind magnetic field ($\theta$ in the schematic in Figure 1e). SBs move in the solar wind frame with a velocity approximately proportional to $\Delta \vec{B}_{SB}$ – the Alfvénicity condition (Kasper et al., 2019). While the dominant magnetic field component inside a SB is typically axial (often close to the T direction), SBs also have transverse components, schematically shown in Figure 1f (Drake et al. 2021; Krasnoselskikh et al. 2020; Larosa et al. 2021). These properties of SBs were well reproduced in numerical simulations of flux ropes in the solar wind (Drake et al., 2021).

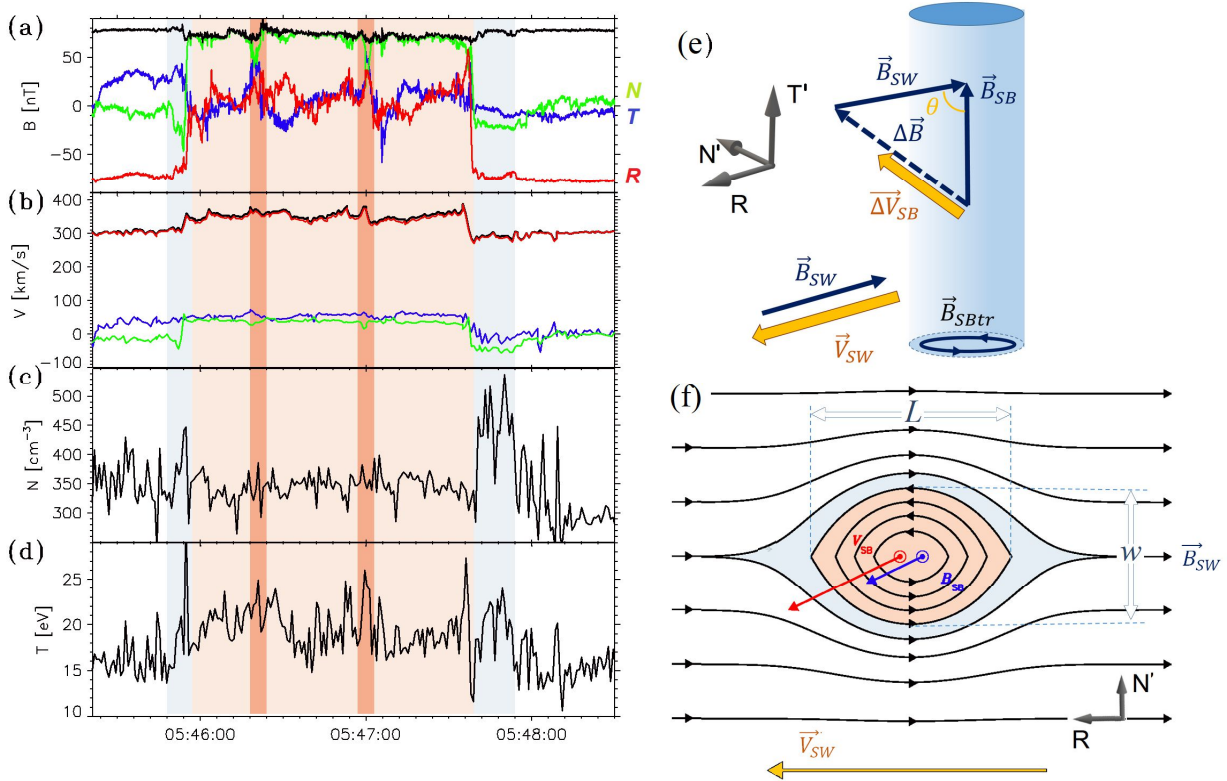

**Figure 1.** A switchback recorded by Parker Solar Probe at 35.7 RS on November 4, 2018 during the first perihelion: (a) - the magnetic field components (in the

RTN system: with the red, blue, and green curves corresponding to $R,T,N$-components). The radial component of the magnetic field exhibits an almost complete rotation inside the switchback and becomes positive (anti-sunward). The magnitude is shown by the black curve; (b) – the proton bulk velocity components (in the RTN system with the same color scheme as in panel (a)) and the absolute value of the bulk velocity (the black curve); (c) - the proton density and (d) the parallel proton temperature. (e) – a flux rope (switchback) schematic with the parameters discussed in the text. (f) – switchback structure in the R-N' plane transverse to the switchback axis (usually close to the R-N plane of the RTN system). The color scheme matches that in panels (a-d) where the boundary region is colored light blue and the core region is light red.

## 3. Analytic analysis of the radial scaling of flux ropes, their aspect ratio and the energetics of merging

A key goal required to establish FRs as possible sources of SBs is to understand the physics basis for their structure that yields observational predictions. Such predictions include the scaling of the size of SBs with radial distance from the sun, their aspect-ratio and their internal magnetic structure, including the large ratio of the axial magnetic field to that defined by the plane of the ambient solar wind. In the following we for simplicity take the solar wind magnetic field to be radial $B_{SW} = B_R$, the axial SB magnetic field $B_{SB}$ to be in the RTN $T$ direction (in a general case the switchback coordinate system RT'N' differs from the RTN system with the $T'$ axis directed along the averaged $\vec{B}_{SB}$, i.e. the SB axis, and the common $R$ axis with the RTN system. However, the following discussion is valid for cases where the angle between $\vec{B}_{SB}$ and $\vec{B}_{SW}$ is in the range of $\pi/2 \pm \pi/6$, which is satisfied for more than 80% of SBs observed by PSP) and the width $w$ of the SB or FR to be in the $N$ direction. (The transverse magnetic field components inside the SB are $B_r$ and $B_N$, and since $|B_R| \approx |B_T|$, $|B_r|$ and $|B_N|$ are much smaller than $|B_R|$ and $|B_T|$.) This coordinate systems can, of course, be generalized so that the ambient solar wind magnetic field lies along the direction of the Parker spiral. We also note that the axial magnetic field might have a component along the radial direction but for simplicity neglect that in the discussion that follows. In the following we presume that SBs are FRs and explore the consequences of this model to understand SB structure.

A key characteristic of FRs concerns the scaling of their area $A$ with radial heliocentric distance $R$. This scaling follows from the conservation of the total axial magnetic flux under the assumption that magnetic reconnection with the ambient solar wind magnetic field is not active. Such an assumption is likely to break down sufficiently far from the sun and is perhaps the reason that SBs are less probable in 1AU observations than closer to the sun. The axial SB magnetic field $B_{SB} = B_T$ nominally scales as $R^{-1}$ due to the expansion of the solar wind in the $N$ direction. However, such a scaling with radius is inconsistent with the balance of magnetic pressure within a flux rope with that of the ambient solar wind radial field $B_R$, which scales as $R^{-2}$. Thus, pressure balance requires that the flux rope area $A$ increase to reduce $B_T$ to match the local $B_R$. Flux conservation then yields $B_T A \sim B_R A \sim R^{-2} A$ so that $A$ scales as $R^2$. In invoking pressure balance we have neglected the magnetic field $B_r$ and $B_N$, the radial and normal magnetic fields of the flux rope, in comparison with $B_T$. This assumption is consistent with most observations. The scaling of $A$ with $R^2$ yields no information on the scaling of the characteristic width $w$ (along $N$) and length $L$ (along $R$) of the flux rope other than $A \sim \pi w L$.

Observations reveal that the aspect ratio $L/w$ of the SBs is large (Horbury et al. 2020; Laker et al. 2021; Mozer et al. 2021). Thus, a fundamental question is what physics leads to such large aspect-ratios? We suggest that it is the weakness of the magnetic field which wraps the flux rope, $B_r$ and $B_N$, compared with $B_R$ and $B_T$, which allows the flux rope to be squashed by the strong solar wind magnetic field. The axial field $B_T$ prevents the compression of the flux rope but provides no restoring force to prevent the flux rope from being squashed to a state in which $B_r/B_N \sim L/w \gg 1$. Note that the magnetic flux $\psi$ that wraps the flux rope is given by $\psi \sim B_N L \sim B_r w$. As the solar wind magnetic field squashes the flux rope, the tension force in the $R$ direction within the flux rope scales like

$$F_R \sim \frac{1}{4\pi} B_N \frac{\partial B_r}{\partial N} \sim \frac{1}{4\pi} B_N \frac{B_r}{w}. \tag{1}$$

The schematic in Figure 1f illustrates the forces involved. This tension force must be balanced by a corresponding tension force $F_N$. In a round FR the balance between these two forces cause the FR to be round. However, the force $F_N \sim \left(\frac{B_r}{4\pi L}\right) B_N$ within the FR is negligible for $L \gg w$ and the restoring force must arise from the weak bending of the solar wind magnetic field due to its distortion by the flux rope. Within the solar wind a weak magnetic field $B_N \sim w B_R/L$ produces the restoring force that limits compression of the FR,

$$F_N \sim \frac{1}{4\pi} B_R \frac{1}{L} \frac{w}{L} B_R. \tag{2}$$

The balance between the two tension forces yields the relation,

$$B_N B_r \sim (w^2/L^2) B_R^2. \tag{3}$$

As the FR aspect ratio changes, $B_N$, $B_r$, $w$ and $L$ change, so an expression for $w$ has to be evaluated at a fixed $A$ and $\psi$, which are invariant as the aspect ratio changes. The resulting expression for $w$ is

$$\frac{w^4}{A^2} \sim \frac{\psi^2}{\pi A B_R^2}. \tag{4}$$

It is convenient to rewrite this relation in terms of more obvious physical parameters as

$$\pi w^2/A \sim B_r^2/B_R^2. \tag{5}$$

This equation reveals that it is the weak magnetic field $B_r$ that wraps the flux rope that allows the FR to be compressed to produce the highly elongated SBs seen in the data. Thus, a fundamental question is why $B_r$ is reduced compared with $B_R$ as FRs propagate outward in the solar wind. It is not a consequence of the simple radial expansion of the solar wind. $B_r$ within the FR has the same scaling properties as $B_R$ since the FR expands in both the $T$ and $N$ directions.

Here we suggest that while FRs that result from interchange reconnection near the solar surface generally have an aspect ratio of order unity, they undergo mergers as they propagate outward in the solar wind and that the merging process reduces the magnetic field $B_r$ below that of the ambient $B_R$. Indeed, it is the reduction of $B_r$ and the associated magnetic energy that facilitates FR merger. The merger of two FRs of similar magnetic flux yields a final FR with increased area $A$ and with a constrained magnetic flux (Fermo et al. 2010). These relations are unchanged when the FR is elongated. However, the final FR aspect ratio and the change in $B_r$ is impacted by the elongation of the FRs. Equation (1) reveals that the FR width $w$ increases by the factor $2^{1/4}$ when the merger of two FRs of equal area and flux merge, which results from the doubling in the area. Since the flux is conserved, $B_r$ is reduced by the factor $(1/2)^{1/4}$. Since $B_r$ dominates $B_N$, the magnetic energy decreases during merger with the energy going to heating the plasma within and around the FR, as has been shown in observations of the coalescence of macroscopic FRs (Drake et al., 2006; Zhou et al. 2014, 2017). Thus, the merger of squashed FRs is energetically favorable and leads to the reduction of the magnetic field that wraps the magnetic flux and increased elongation of FRs.

This result suggests that FRs should be increasingly elongated with radial distance from the sun. Further, the reduction of $B_r$ and also $B_N$ within the FR means that the axial magnetic field of FRs dominates that of the other components. This explains one of the key features of SBs observations, the sharp rotation of the magnetic field in the solar wind into the axial direction upon entry into a SB.

## 4. Dynamics of the process of SBs merging: the numerical results

Drake et al. (2021) presented a model of SB generation by interchange reconnection between open and closed flux in the low corona that created flux ropes that ejected them with high velocity outward in the solar wind. The structures have a strong axial magnetic field wrapped by magnetic flux and exhibit the characteristic internal rotation of the radial magnetic field. The dynamics of this system reproduced well the magnetic structure of SBs seen in PSP data and also indicated the tendency of flux ropes to merge. We focus here on the details of this merging process to identify the observational features and the consequences for SB structure. Interchange reconnection favors the production of a series of flux ropes that have similar axial magnetic fields. We perform numerical simulations with the PIC code p3d (Zeiler et al. 2002) using a setup similar to that presented in Drake et al. (2021). The initial field configuration consists of a straight background magnetic field $B_0$ directed along the $R$-axis (we adapt the numerical system coordinates to the RTN system in the solar wind), a weak initial reversed magnetic field ($\sim 0.2B_0$) and the guide field $B_T$ in the region where the initial radial field reverses to be of order $B_R$, so, that the total magnetic field magnitude is a constant across the region of reversed flux. The initial plasma density and temperature are uniform. The simulation results are presented in normalized units: magnetic field to $B_0$, time to $\Omega_i^{-1}$ and distance to the ion inertial length $d_i$. The computational domain is given by $L_x \times L_y = 40.96 d_i \times 40.96 d_i$ with the grid spacing given by $\delta x = \delta y = 0.05 d_i$, and 100 particles per cell. The reversed magnetic field, which drives reconnection and eventually produces the magnetic field $B_R$ and $B_N$ that wraps the flux rope, is weak compared with $B_T$ as in SB observations from PSP data.

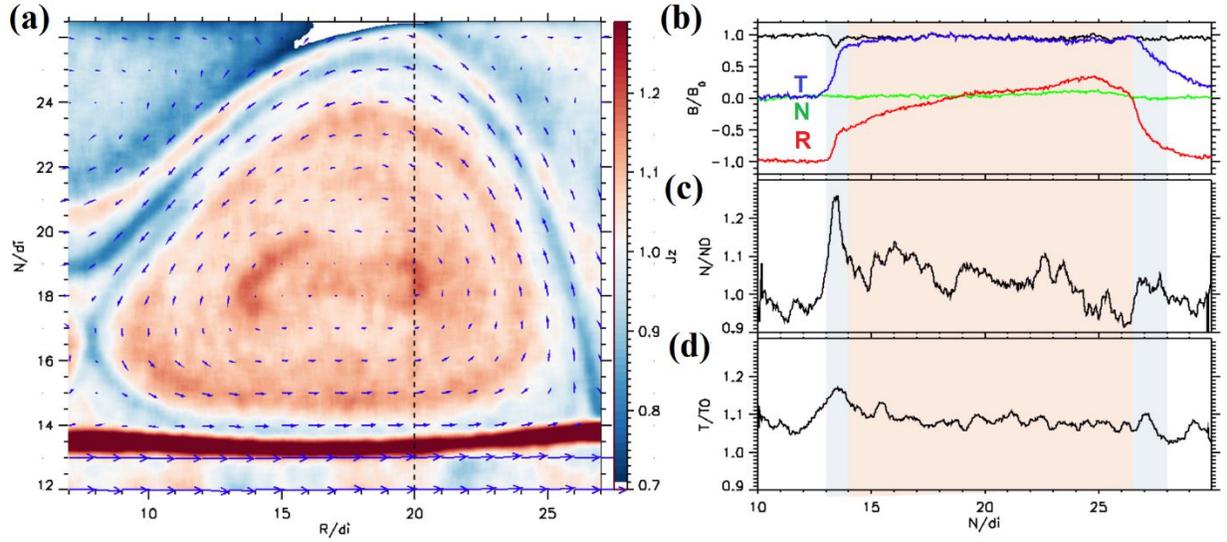

**Figure 2.** The SB structure from the PIC simulation ($\Omega_i t = 270$) in the R-N plane (similar cut as in the schematic in Figure 1f): (a) - the transverse magnetic field (arrows) and the axial current $J_T$; (b) – the structure of the magnetic field (the $B_R$, $B_T$, and $B_N$ components are shown by the red, green, and blue curves respectively); (c) and (d) – the plasma density and temperature. The cuts are along the N direction through the center of the flux rope.

Reconnection started from noise leads to generation of many flux ropes, which then merge. The magnetic field structure transverse to the SB axis is shown in Figure 2a superimposed over the axial electron current. It reveals the characteristic magnetic island structure with wrapped magnetic field components, $B_R$ and $B_N$. Shown in Figure 2 are the magnetic field components (in b), the density (in c) and the temperature (in d). The magnetic field configuration and the plasma parameters are in a good agreement with SB structure obtained from PSP measurements: sharp rotation of the magnetic field direction at the SB boundaries, almost constant magnetic field magnitude and plasma density inside with localized magnetic dips (Farrell et al., 2020; Agapitov et al., 2020) and density enhancements at the boundaries (Farrell et al., 2020; Larosa et al., 2021).

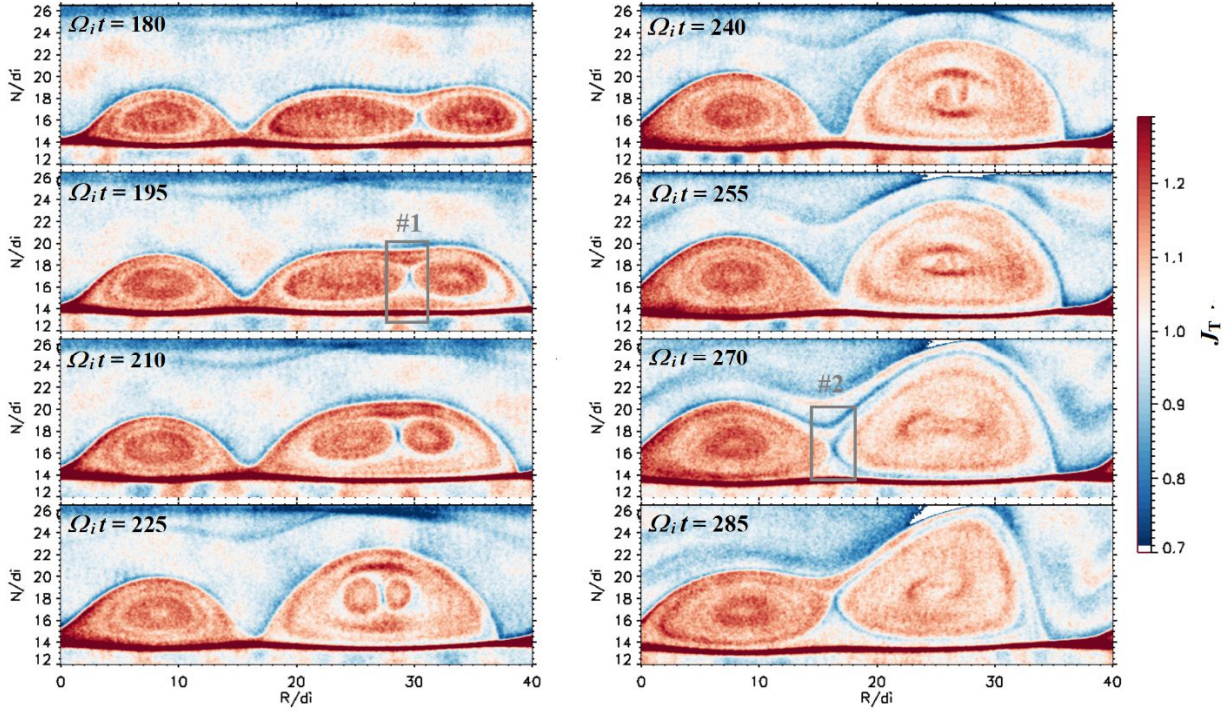

**Figure 3.** The dynamics of the flux ropes merging shown in the out-of-plane current $J_T$ in the R-N plane.

The time series of three merging FRs (from left to right, FR1, FR2 and FR3) from the simulation is shown in Figure 3. First, FR2 and FR3 merge into a single larger structure FR2-3 with lower internal transverse magnetic field (Drake et al. 2013). Later, at $\Omega_i t = 225$, FR1 approaches FR2-3 and at around $\Omega_i t = 270$ merging of FR1 and FR2-3 starts.

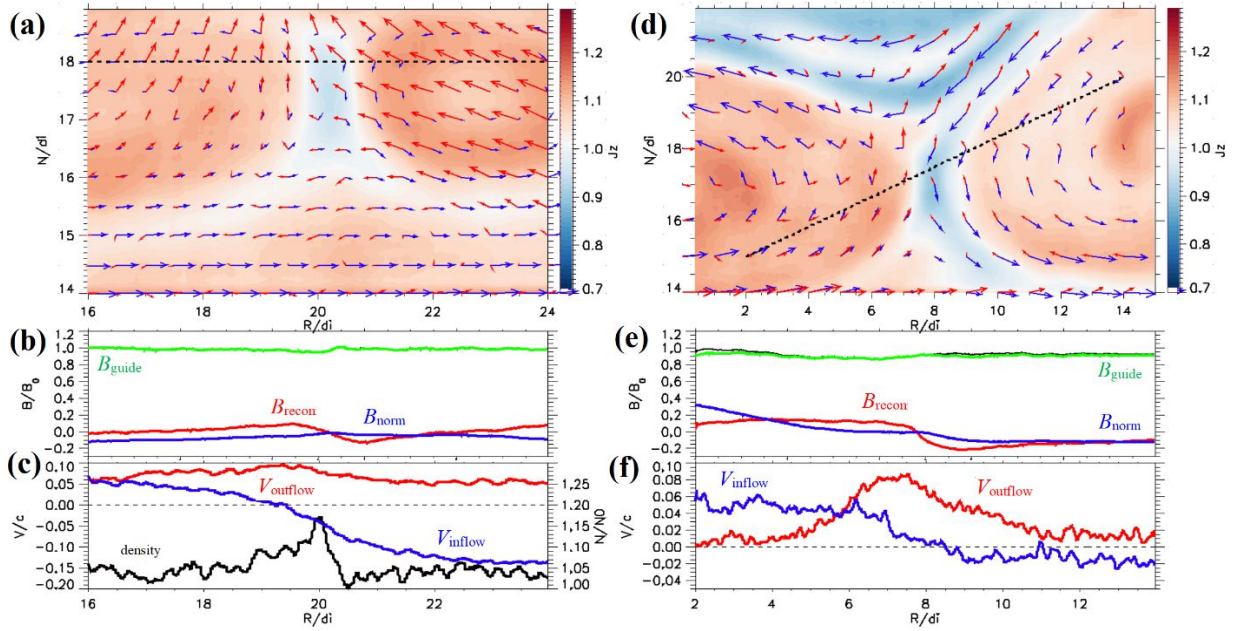

**Figure 4.** The reconnection regions (marked in Figure 3 with the grey boxes) during the merging of FR2 and FR3 (in panels a-c) and FR2-3 and FR1 (panels d-f). Panel (a) shows the out-of-plane electron current $J_{eT}$, and the structure of the magnetic field (blue arrows) and plasma flow (red arrows) in the reconnection region #1 (FR2 and FR3 merging). The data along the black dashed line are shown in panels (b) and (c): in panel (b) the components of magnetic field in the current sheet coordinate system (the guide field is green, the normal component is blue, and the reconnecting component is red); and in panel (c) – the in-plane components of plasma flow velocity (the blue curve is the inflow velocity – the normal component to the current sheet; the red curve is the outflow velocity) with the plasma density shown by the black curve with the scale in the right. The right panels (d,e,f) show the merger of FR1 and FR2-3.

The reconnection regions (zoomed images of boxes #1 and #2 from Figure 3) shown in Figure 4 present the details of the magnetic field and velocity structure during the merging of the FRs. A current sheet develops between the merging islands. It is predominantly in the N-T plane and has a width about the proton inertial length in both cases in Figure 4. Across the current sheet $B_N$ changes sign. The inflow velocity $V_{inflow}$ normal to the reconnection current sheet plane (the blue curves) reveals that plasma flows toward the reconnecting current sheet with velocity about 0.5 of the Alfvén velocity based on the reconnecting magnetic field magnitude ($V_{Ar}$) in the second case (Figure 4d-f). The plasma outflow is directed mostly along the N-axis with velocity about 0.8-

0.9 $V_{Ar}$. In the first case (Figure 4a-c) the inflow and outflow velocities are similar but the motion of the right flux rope provides an additional velocity ~0.5 $V_{Ar}$ that leads to an increase of the negative values to about $V_{Ar}$. The first case (Figure 4a-c) does not show a significant outflow so that there is no change of in the sign of the correlation of $B_{recon}$ and $V_{outflow}$ during the current sheet crossing.

5. Parker Solar Probe observations of switchback structure

The features of merged flux ropes seen in the simulations, including localized current layers and localized density and temperature enhancements, are often seen in the switchback structures observed in the solar wind by PSP. An important question is therefore whether there is evidence for merging in the observational data. An example is presented in Figure 5 showing two switchbacks (highlighted by red in Figure 5) approaching each other and driving a density enhancement between them (highlighted by blue). The second switchback has a complex inner structure of magnetic field and plasma velocity perturbations: it consists of four regions with three transition regions – current sheets (CS) marked by deep red. Based on the structure of flux ropes from the simulation we suggest that the second switchback consists of four or five flux ropes. This scenario is supported by the structure of the plasma density and temperature, which have sharp, localized enhancements around the CS's. The structure of the three CS's (highlighted by red in Figure 5) is shown in three expanded views in Figure 6.

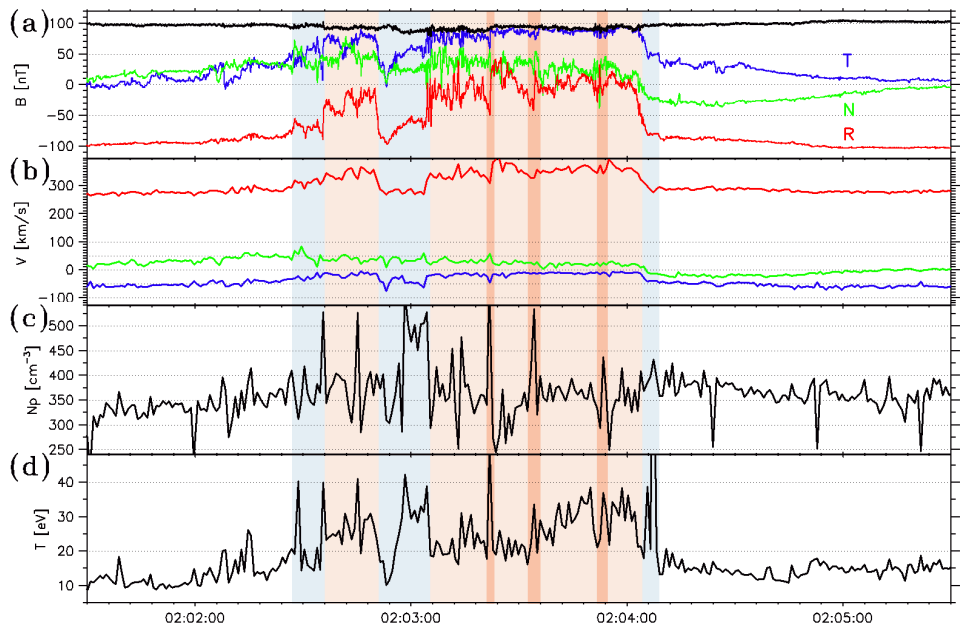

**Figure 5.** Two switchbacks (highlighted by light-red) recorded on November 6, 2018 by Parker Solar Probes. The panels from top to bottom present the magnetic field in the RTN coordinate system - (a) and the proton bulk velocity - (b); the proton density - (c) and the parallel proton temperature - (d). The deep red regions highlight boundaries between distinct regions of the second switchback.

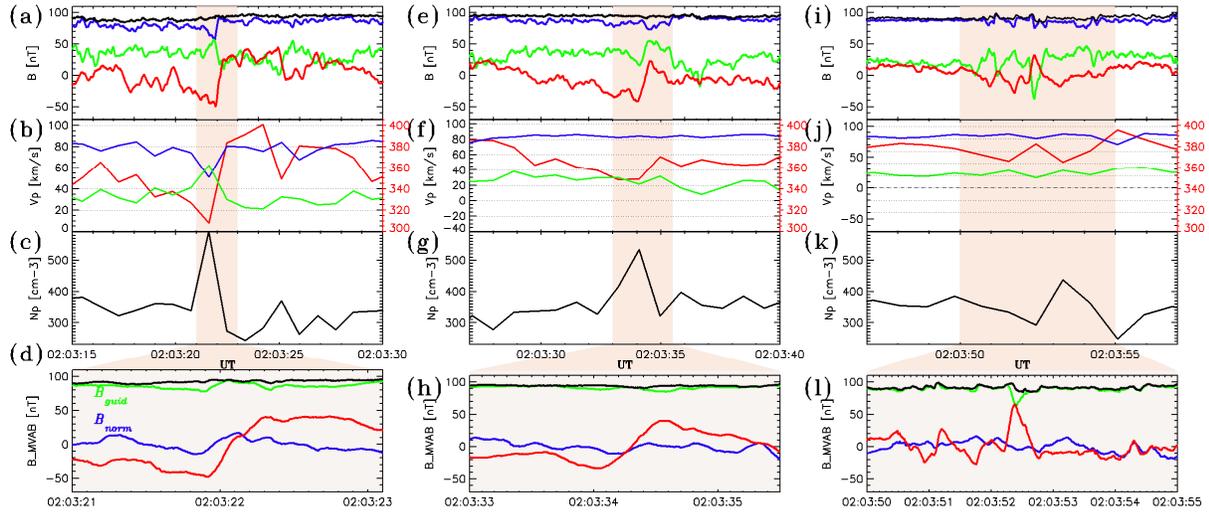

**Figure 6.** The zoomed intervals highlighted in Figure 5. CS#1: (a) – the magnetic field components and magnitude; (b) – plasma bulk velocity components; (c) – proton density; (d) – the magnetic field component in the current sheet frame (the interval highlighted in panel (b)). The second and third intervals (CS#2 and CS#3) highlighted in Figure 5 are presented in panels (e-h) and (i-l) respectively.

The first interval (CS#1: 2:03:15-2:03:30) reveals signatures that would seem to suggest that merging of two flux ropes recently terminated. The transverse-to-the-flux-rope-axis magnetic field (red) reverses sign across a well-defined current layer that produces a magnetic field jump of ±(50±5) nT. The guide field (blue) is 90±5 nT, i.e. ~2 times greater than the maximal value of the transverse field. The CS thickness is 12±3 km (the proton inertial length is 12±1 km). The structure of the magnetic field for CS#1 is similar to that of the reconnecting current sheet between the two merging flux ropes from the simulation in Figures 4a-c. However, the velocities for this interval shown in Figure 6b do not reveal the characteristic Alfvenic reconnection outflow centered on the current layer, which persists even in the case of reconnection with a strong guide field (Drake et al. 2020; Gosling & Phan 2013; Phan et al.

2020). Rather, the flows exhibit the typical Alfvénic relation between velocity and magnetic that has been documented in earlier PSP data (Kasper et al 2019, Phan et al 2020). Thus, in spite of the intense current layer seen in this interval, the velocity data does not support the idea that reconnection is ongoing. Nevertheless, the presence of strong current layers and other signatures that are normally attributed to active reconnection requires explanation.

A limitation of the reconnection and merging simulations presented in Section 4 was the absence of the characteristic Alfvenic flows present in the solar wind. These Alfvenic flows might prevent flux rope merging since it is known that sheared flows across a current layer can prevent reconnection (Chen et al. 1997; Cowley & Owen 1989). We have initiated a simulation study of flux rope merger that includes parallel flows with $\boldsymbol{V} = \alpha \boldsymbol{V}_A \boldsymbol{b}$ in the initial condition. The parameter $\alpha$ is a constant that typically ranges between zero and one, the latter corresponding to fully Alfvenic flows. During periods in the solar wind when flows are Alfvenic in character, the parameter $\alpha$ is typically of order unity or less. To study flux rope merger, we initialize the system with two equal-sized, cylindrical flux ropes with a strong ambient guide field that has a magnitude that is twice the peak in-plane magnetic field. The initial flux rope equilibrium is the same as that reported previously (Drake et al. 2021). The two flux ropes are overlapped slightly to initiate reconnection. Here we show an example from one of the simulations to illustrate the qualitative behavior of FR merger and its relation to the PSP observations. Further details will be presented in a more complete paper (Swisdak et al 2021). Simulation data for the parameter $\alpha = 0.75$ is shown in Figure 7. In (a,b) and (c,d) are the out-of-plane current with overlying magnetic field lines at two times. In (a) reconnection is well-developed and a strong current layer has developed between the two FRs. At this time strong outflows from the magnetic x-line have developed as shown in (b). The outflows are nearly centered on the current layer as expected in a traditional reconnection outflow. Note the downflow on the right side of the FR and the upflow on the left that correspond to plasma circulation within each of the FRs. At late time in (c) and (d) magnetic reconnection and FR merging has ended. Nevertheless, a strong current layer remains: the cut along the black line in panel (c) reveals the reversal of By with almost constant $B_x$ and $B_z$ (the guide field) in panel (e); the reversal of By with the corresponding component of plasma flow velocity in panel (f) indicate the evidence that along a cut across the current layer, the direction of the vertical flow reverses along with the reversal in the corresponding vertical magnetic field (as seen in the observations in Figure 6a,b and Figure 6e,f) and as is required to maintain the Alfvénicity of the flows); the enhancement of plasma density and proton temperature in the current layer are shown in panel (g) and are also consistent with the observations (Figure 7g). At this time the FRs begun to rotate around each other with the FR on the right moving up and the FR on the left moving down. This is evident from the displacement of the FRs in (c) as well as the flows in (d). Thus, the late time structure of the magnetic field and flows are qualitatively consistent with the observational data in Figure 6.

The saturation of the FR merging is a consequence of energy transfer from the released magnetic energy into the plasma flow circulating on the reconnecting field line. As the reconnected field line shortens, the parallel streaming velocity increases, increasing the Alfvénicity inside the FRs because of the invariance of the action $V_{\parallel}L$, with L the field line length. When the Alfvénicity approaches unity, reconnection is energetically unfavorable. This has important implications for understanding the measured Alfvénicity in the PSP observations (Swisdak et al. 2021). The increase of the Alfvénicity as a consequence of the merging suggests the possible connection of the Alfvénicity with the proton temperature inside the FRs since merger increases both the Alfvénicity and the plasma temperature. The variation in several plasma parameters inside the different flux ropes is displayed in Figure 8: the magnetic fields $B$ and $B_T$ (Figure 8a), the radial magnetic field $B_R$ (Figure 8b), the Alfvénicity (Figure 8c), and plasma temperature (Figure 8d). The flux ropes are separated by current sheets and the FRs with higher proton temperature inside have higher levels of Alfvénicity (Figure 8e). This suggests that the increase in both parameters is likely the result of merging but that merging has now ended, leaving the remnant current layers separating distinct flux ropes as shown in Figure 7.

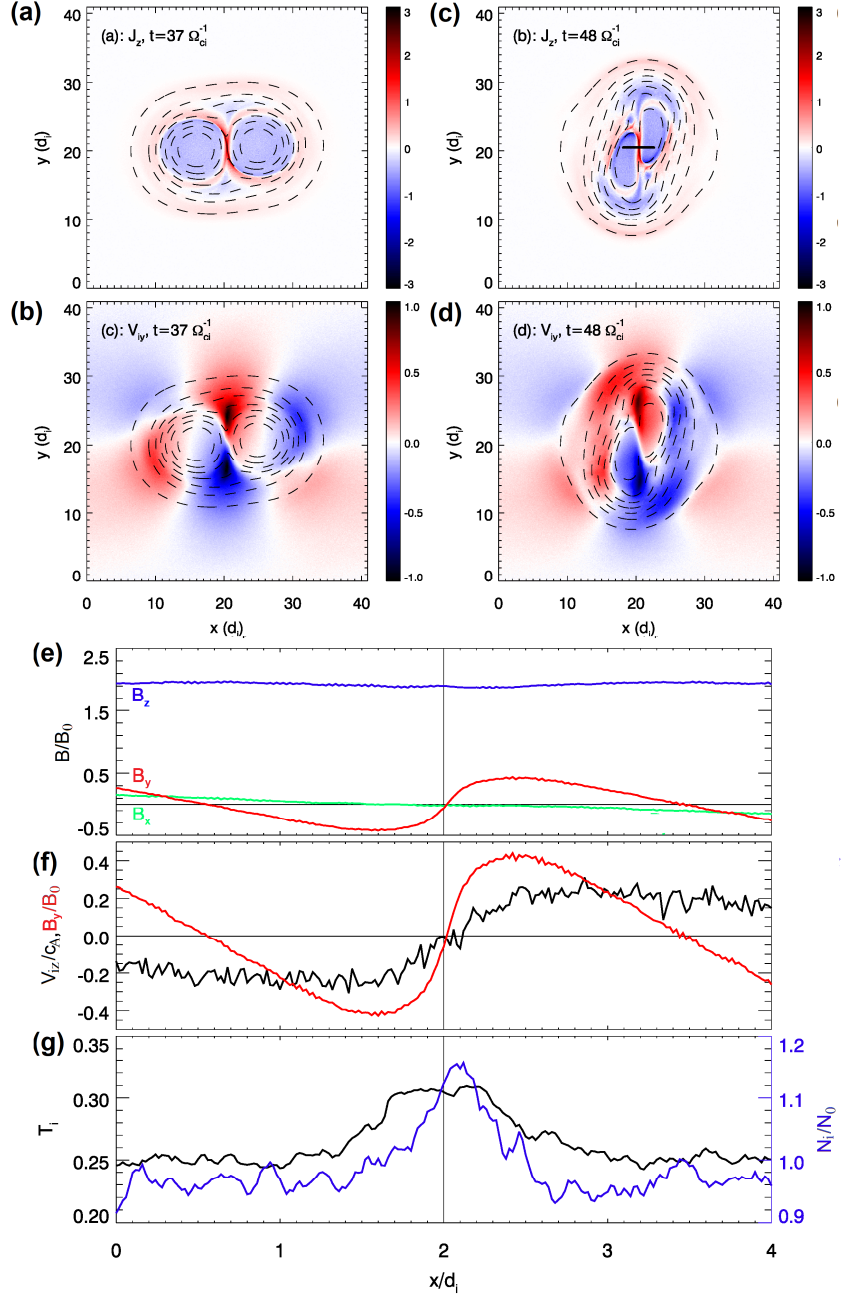

**Figure 7.** The results of a simulation of flux rope merger with initial Alfvenic flow. Out-of-plane current and magnetic field lines during merging in (a) and after merging ends in (c). Vertical flows $V_y$ in (b) and (d) corresponding to the times in (a) and (c). The data along the black line in panel (c) is shown in panels (e-g): (e) – the magnetic field; (f) – the reversing magnetic field component ($B_y$ – the red curve) and the corresponding component of the plasma flow velocity ($V_y$ – the blue curve); (g) – the plasma density (the blue curve) and temperature (the black curve).

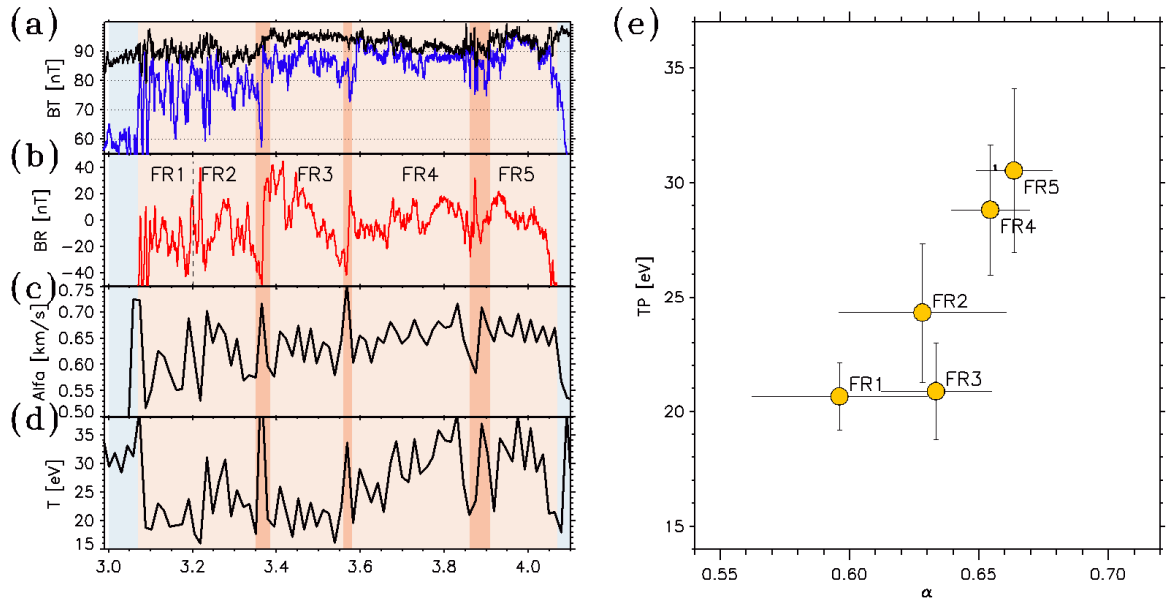

**Figure 8.** The structure of the switchback from Figure 5: (a) – the T-component of magnetic field (the blue curve); (b) – the radial component of magnetic field (the R-component) indicating the magnetic island structure of the SB components; (c) – the Alfvénicity α inside the SB; (d) – the radial temperature inside the SB; (e) – the dependence of the proton temperature on the Alfvénicity for the structure components (individual flux ropes) composing the SB.

The second current sheet (2:03:33-2:03:40, CS#2) presents a crossing of the current sheet with a magnetic field change of ±(35±5) nT (the guide field of 90±5 nT is similar to CS#1) and the velocity following the changes of the reconnecting magnetic field. The current layer thickness is 16±5 km (the proton inertial length is 12±1 km) with a similar density enhancement as in CS#1. The third interval (2:03:47-2:03:57, CS#3) does not reveal a strong current sheet and reversed magnetic field. The peak in the density suggests that this boundary could correspond to a post-merging configuration. Thus, CSs#1,2,3 are possibly examples of the current sheets resulting from merging of flux ropes in the solar wind and conserved in time due to increased (in the process of merging) Alfvénicity.

Due to significant elongation along the radial direction switchbacks most probably merge along their long dimension – the current sheets in CSs#1,2,3 have normals directed predominantly along *N*-axis (the schematic of the system geometry is shown in Figure 9).

The structure of perturbations suggest that the large switchback in Figure 5 is the result of the partial merging of four (or five) flux ropes with similar parameters (magnetic field magnitude and direction, plasma density) that probably originated from the same source.

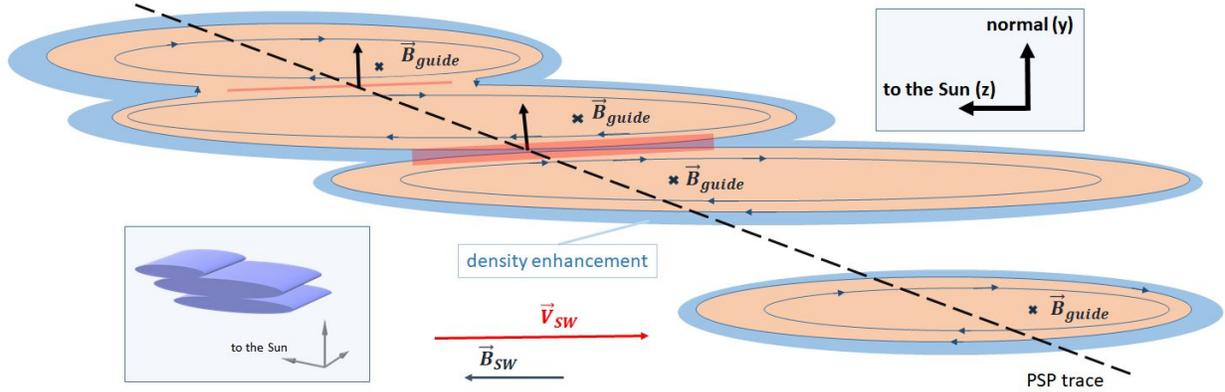

**Figure 9**. The schematic of the structure of switchbacks from Figure 5 in the R-N plane. The scales are arbitrary. The red arrows show the outflow at the edges of reconnection current sheet. The black arrows show the direction of a normal to the current sheet at the point of crossing by the spacecraft.

6. Discussion and Conclusions

Drake et al. (2021) showed that flux ropes can form in the low corona through interchange reconnection and can be injected into the solar wind. Flux ropes in reconnecting current sheets are generated at small spatial scales as current sheets narrow and reconnection develops (Bhattacharjee et al. 2009; Biskamp 1986, 1986; Cassak et al. 2009; Drake et al. 2006). We have shown here that

1. While flux ropes that result from interchange reconnection near the solar surface are generally likely to form with an aspect ratio of order unity and comparable axial and transverse magnetic field, they undergo geometrical changes while propagating outward in the solar wind, tending to significant elongation along the background magnetic field, and interact with each other through merging.
2. Merging occurs through the slow reconnection of the weak magnetic field that wraps the stronger axial magnetic field and, thus, reduces the strength of the wrapping magnetic field and heats the plasma inside the structure. Merging of flux ropes is energetically favorable and increases the axial plasma flow speed leading to increased Alfvénicity of the structure.

3. When the Alfvénicity approaches unity ($\Delta\vec{B}_{SB}/(4\pi m_p)^{1/2} \approx \Delta\vec{V}_{SB}$) merging is becomes energetically unfavorable, and thus the saturation of flux rope merging is a consequence of energy transfer from the reconnected magnetic field into the plasma flow and thermal energy. When the flux rope Alfvénicity becomes significant flux rope merger saturates before it is complete, which leads to remnant current sheets with magnetic and velocity characteristics consistent with PSP observations. Thus, the multi-flux rope structure with the conserved in time remnant current sheaths that characterizes many SW's is a consequence of incomplete flux rope merger.
4. This has important implications for understanding the measured Alfvénicity in the PSP observations. The strength of the wrapping magnetic field (decreasing through flux rope merging) controls the elongation of flux ropes: a weaker wrapping magnetic field allows the ambient solar wind magnetic field to squash and elongate the flux ropes (and therefore switchbacks). Thus, switchbacks become increasingly elongated along the solar wind magnetic field with radial distance from the sun. The result is that the switchbacks evolve to a state with a weak magnetic field that wraps the switchback compared to it's axial field. Therefore, a sharp rotation of the magnetic field is observed at switchback boundaries.
5. The reduction of the magnetic field that wraps the flux rope during merging might be responsible for the observed plasma temperature enhancement inside switchbacks. Thus, the signature of flux ropes mergers can be the relation of plasma temperature and Alfvénicity level inside a switchback since both increase during merging. The signatures of switchback merging similar to those obtained in the numerical modeling are often seen in PSP observations of switchbacks at 20-50 solar radii. This suggests that merging of flux ropes is a significant part of the evolution of switchbacks from the flux ropes generated in the low corona to the magnetic structures observed by PSP.

## Appendix A1

Magnetic energy versus time during the merger of flux ropes FR2 and FR3 is presented in Figure A1: the total magnetic energy ($B_{SB}^2/8\pi$) - light red in Figure A1a; transverse (wrapped) magnetic field energy (($B_R^2 + B_N^2)/8\pi$) - dark red in Figure A1a. Plotted in Figure A1b is the ratio of transverse magnetic field $B_R^2 + B_N^2$ divided by the total magnetic field $B_{SB}^2$ averaged over the R-N cross-section area of the combined FR2-3. FR2 and FR3 are merging (the interval of merging is highlighted by yellow), and that significantly changes their magnetic field structure, leading to a fast decrease of the transverse magnetic field from 0.04 to 0.02. Thus, merging of flux ropes leads to fast decay of the transverse (wrapped) components of magnetic field.

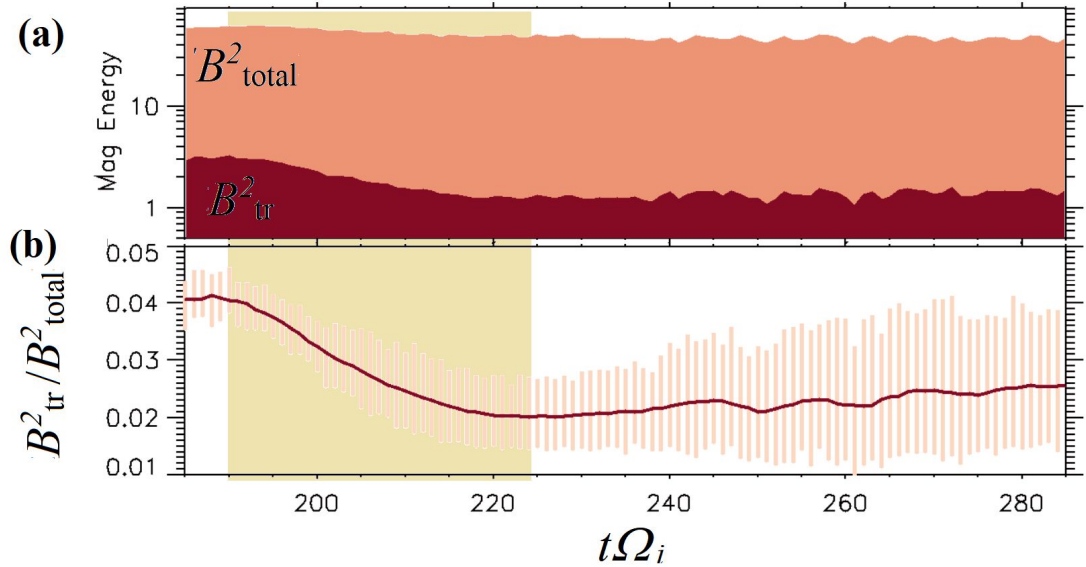

**Figure A1**. Dynamics of the merger of flux ropes FR2 and FR3. Magnetic energy (panel a - the total energy is marked with light red and the transverse energy is marked with dark red) and (b) -the contribution from the transversal (wrapped) magnetic field (averaged over the entire flux rope cross-section area) in FR2-3 from Figure 3. The error bars represent the values averaged over different crossings of FB2-3 by a virtual spacecraft. The interval of FB2 and FB3 merging is highlighted with yellow.

ACKNOWLEDGEMENTS

OVA and JFD were supported by NASA grant 80NNSC19K0848; OA was partially supported by NSF grant number 1914670 and NASA Living with a Star (LWS) program (contract 80NSSC20K0218).

# 5. References

Agapitov, O. V., Wit, T. D. de, Mozer, F. S., et al. 2020, ApJL, 891 (American Astronomical Society), L20
Akhavan-Tafti, M., Kasper, J., Huang, J., & Bale, S. 2021, A&A (EDP Sciences), https://www.aanda.org/articles/aa/abs/forth/aa39508-20/aa39508-20.html
Bale, S. D., Badman, S. T., Bonnell, J. W., et al. 2019, Nature, 576 (Nature Publishing Group), 237
Bale, S. D., Goetz, K., Harvey, P. R., et al. 2016, Space Sci Rev, 204, 49

Balogh, A., Forsyth, R. J., Lucek, E. A., Horbury, T. S., & Smith, E. J. 1999, Geophysical Research Letters, 26, 631

Bhattacharjee, A., Huang, Y.-M., Yang, H., & Rogers, B. 2009, Physics of Plasmas, 16 (American Institute of Physics), 112102

Biskamp, D. 1986, The Physics of Fluids, 29 (American Institute of Physics), 1520

Borovsky, J. E. 2016, Journal of Geophysical Research: Space Physics, 121, 5055

Case, A. W., Kasper, J. C., Stevens, M. L., et al. 2020, ApJS, 246, 43

Cassak, P. A., Shay, M. A., & Drake, J. F. 2009, Physics of Plasmas, 16 (American Institute of Physics), 120702

Chen, Q., Otto, A., & Lee, L. C. 1997, Journal of Geophysical Research: Space Physics, 102, 151

Chen, Y., & Hu, Q. 2020, ApJ, 894, 25

Chen, Y., Hu, Q., Zhao, L., et al. 2020, ApJ, 903, 76

Chen, Y., Hu, Q., Zhao, L., Kasper, J. C., & Huang, J. 2021, ApJ, 914, 108

Chen, Y., & Hu, Q. 2021, arXiv:211109261 [astro-ph, physics:physics], http://arxiv.org/abs/2111.09261

Cowley, S. W. H., & Owen, C. J. 1989, Planetary and Space Science, 37, 1461

Dahlburg, R. B., Boncinelli, P., & Einaudi, G. 1997, Physics of Plasmas, 4 (American Institute of Physics), 1213

Drake, J. F., Agapitov, A., Swisdak, M., et al. 2020, A&A (EDP Sciences), https://www.aanda.org/articles/aa/abs/forth/aa39432-20/aa39432-20.html

Drake, J. F., Agapitov, O., Swisdak, M., et al. 2021, A&A, 650 (EDP Sciences), A2

Drake, J. F., Swisdak, M., Che, H., & Shay, M. A. 2006, \nat, 443, 553

Dudok de Wit, T., Krasnoselskikh, V. V., Bale, S. D., et al. 2020, ApJS, 246, 39

Farrell, W. M., MacDowall, R. J., Gruesbeck, J. R., Bale, S. D., & Kasper, J. C. 2020, ApJS, 249 (American Astronomical Society), 28

Fermo, R. L., Drake, J. F., & Swisdak, M. 2010, Physics of Plasmas, 17 (American Institute of Physics), 010702

Fisk, L. A., & Kasper, J. C. 2020, ApJL, 894 (American Astronomical Society), L4

Fox, N. J., Velli, M. C., Bale, S. D., et al. 2016, Space Sci Rev, 204, 7

Froment, C., Krasnoselskikh, V., Wit, T. D. de, et al. 2021, A&A (EDP Sciences), https://www.aanda.org/articles/aa/abs/forth/aa39806-20/aa39806-20.html

Gosling, J. T., & Phan, T. D. 2013, ApJL, 763 (American Astronomical Society), L39

He, J., Zhu, X., Yang, L., et al. 2020, arXiv:200909254 [astro-ph, physics:physics], http://arxiv.org/abs/2009.09254

Horbury, T. S., Matteini, L., & Stansby, D. 2018, Monthly Notices of the Royal Astronomical Society, 478, 1980

Horbury, T. S., Woolley, T., Laker, R., et al. 2020, ApJS, 246, 45

Jannet, G., Wit, T. D. de, Krasnoselskikh, V., et al. 2021, Journal of Geophysical Research: Space Physics, 126, e2020JA028543

Kahler, S. W., Crocker, N. U., & Gosling, J. T. 1996, Journal of Geophysical Research: Space Physics, 101, 24373

Kasper, J. C., Abiad, R., Austin, G., et al. 2016, Space Sci Rev, 204, 131

Kasper, J. C., Bale, S. D., Belcher, J. W., et al. 2019, Nature, 576 (Nature Publishing Group), 228

Krasnoselskikh, V., Larosa, A., Agapitov, O., et al. 2020, arXiv:200305409 [astro-ph, physics:physics], http://arxiv.org/abs/2003.05409

Laker, R., Horbury, T. S., Bale, S. D., et al. 2021, A&A, 650 (EDP Sciences), A1

Landi, S., Hellinger, P., & Velli, M. 2006, Geophysical Research Letters, 33, https://agupubs.onlinelibrary.wiley.com/doi/abs/10.1029/2006GL026308
Larosa, A., Krasnoselskikh, V., Wit, T. D. de, Agapitov, O., & Froment, C. 2021, A&A (EDP Sciences), https://www.aanda.org/articles/aa/abs/forth/aa39442-20/aa39442-20.html
Lee, L. C., & Fu, Z. F. 1985, Geophysical Research Letters, 12, 105
Macneil, A. R., Owens, M. J., Wicks, R. T., & Lockwood, M. 2020, Monthly Notices of the Royal Astronomical Society, https://doi.org/10.1093/mnras/staa3983
Martinović, M. M., Klein, K. G., Kasper, J. C., et al. 2020, ApJS, 246, 30
Mozer, F. S., Agapitov, O. V., Bale, S. D., et al. 2020, ApJS, 246 (American Astronomical Society), 68
Mozer, F. S., Bale, S., Bonnell, J., et al. 2021, arXiv:210507601 [astro-ph, physics:physics], http://arxiv.org/abs/2105.07601
Neugebauer, M., & Goldstein, B. E. 2013, AIP Conference Proceedings, 1539 (American Institute of Physics), 46
Odstrcil, D., Vandas, M., Pizzo, V. J., & MacNeice, P. 2003, 679, 699
Oka, M., T. D. Phan, S. Krucker, M. Fujimoto, and I. Shinohara, Astrophys. J., 714, 915-926, doi:10.1088/0004-637x/714/1/915 (2010).
Pritchett P. L., Phys. Plasmas 14, 052102 (2007).
Phan, T. D., Bale, S. D., Eastwood, J. P., et al. 2020, ApJS, 246 (American Astronomical Society), 34
Phan, T. D., Gosling, J. T., Paschmann, G., et al. 2010, ApJL, 719 (American Astronomical Society), L199
Phan, T. D., Paschmann, G., Gosling, J. T., et al. 2013, Geophysical Research Letters, 40, 11
Ruffolo, D., Matthaeus, W. H., Chhiber, R., et al. 2020, ApJ, 902, 94
Schwadron, N. A., & McComas, D. J. 2021, arXiv:210203696 [astro-ph, physics:physics], http://arxiv.org/abs/2102.03696
Russell, C. T., & Elphic, R. C. 1978, Space Sci Rev, 22, 681
Shoda, M., Chandran, B. D. G., & Cranmer, S. R. 2021, arXiv e-prints, arXiv:2101.09529
Slavin, J. A., Lepping, R. P., Gjerloev, J., et al. 2003, Journal of Geophysical Research: Space Physics, 108, SMP 10
Song, H.297 Q., Y. Chen, G. Li, and X. L. Kong, Phys. Rev. X, 2, 021015 (2012).
Sterling, A. C., & Moore, R. L. 2020, ApJ, 896, L18
Swisdak, M., Opher, M., Drake, J. F., & Bibi, F. A. 2010, ApJ, 710 (American Astronomical Society), 1769
Tenerani, A., Velli, M., Matteini, L., et al. 2020, ApJS, 246, 32
Whittlesey, P. L., Larson, D. E., Kasper, J. C., et al. 2020, ApJS, 246 (American Astronomical Society), 74
Woodham, L. D., Horbury, T. S., Matteini, L., et al. 2020, arXiv:201010379 [astro-ph, physics:physics], http://arxiv.org/abs/2010.10379
Yamauchi, Y., Suess, S. T., Steinberg, J. T., & Sakurai, T. 2004, Journal of Geophysical Research: Space Physics, 109, https://agupubs.onlinelibrary.wiley.com/doi/abs/10.1029/2003JA010274
Zank, G. P., Nakanotani, M., Zhao, L.-L., Adhikari, L., & Kasper, J. 2020, ApJ, 903 (American Astronomical Society), 1
Zeiler, A., Biskamp, D., Drake, J. F., et al. 2002, Journal of Geophysical Research: Space Physics, 107, SMP 6
Zhao, L.-L., Zank, G. P., Adhikari, L., et al., 2020, *ApJS*, *246*(2), 26.
Zhou, M., Berchem, J., Walker, R. J., et al. 2017, Phys Rev Lett, 119 (American Physical Society), 055101
Zhou, M., Pang, Y., Deng, X., Huang, S., & Lai, X. 2014, Journal of Geophysical Research: Space Physics, 119, 6177